\documentclass{PoS}
\usepackage{amsmath}
\title{Diffractive vector meson production at HERA using holographic AdS/QCD wavefunctions}

\ShortTitle{Diffractive vector meson production at HERA using holographic AdS/QCD wavefunctions}

\author{Jeff FORSHAW\\
        University of Manchester\\
        E-mail: \email{jeff.forshaw@hep.manchester.ac.uk}}

\author{\speaker{Ruben SANDAPEN}%
         \thanks{RS thanks the organisers for their invitation as well as Universit\'e de  Moncton $\&$ Mount Allison University for funding.}\\
        Universit\'e de  Moncton $\&$ Mount Allison University \\
        E-mail: \email{ruben.sandapen@umoncton.ca}}

\abstract{We demonstrate another success of the AdS/QCD correspondence by showing \cite{Forshaw:2012im,Forshaw:2012mb} that an AdS/QCD holographic light-front wavefunction for the $\rho$ meson generates predictions for the cross-sections of diffractive $\rho$ production that are in agreement with data collected
at the HERA electron-proton collider \cite{Chekanov:2007zr,Collaboration:2009xp}. }

\FullConference{XXI International Workshop on Deep-Inelastic Scattering and Related Subjects\\
                 22-26 April, 2013\\
                 Marseilles, France}
\usepackage{bm}

\def\d{\mathrm{d}}
\begin{document}

\section{Introduction}
The AdS/QCD correspondence \cite{Erdmenger:2007cm,CasalderreySolana:2011us,Costa:2012fw,deTeramond:2012rt} refers to the connection between QCD in physical spacetime and string theory in a higher dimensional anti-de Sitter (AdS) space. The precise nature of this connection has not yet been elucidated but there is growing evidence, to which we add here, that there exists such a connection. One particular realization of this connection is light-front holography \cite{deTeramond:2008ht} proposed by Brodsky and de T\'eramond. In light-front holography, the confining QCD potential at equal light-front time between a quark and antiquark in a meson is determined by the profile of the dilaton field which breaks conformal invariance  of the higher dimensional AdS space in which strings propagate. 

In a semi-classical approximation to light-front QCD, Brodsky and de T\'eramond derived a Schroedinger-like equation for mesons: 
\begin{equation}
\left(-\frac{\d^2}{\d \zeta^2} - \frac{1-4L^2}{4 \zeta^2} + U(\zeta) \right) \Phi (\zeta)=M^2 \Phi (\zeta) \;,
\label{LFeigenvalue}
\end{equation}
where $\zeta=\sqrt{x(1-x)} r$ is the transverse separation between the quark and antiquark at equal light-front time \footnote{$x$ is the fraction of light-front momentum carried by the quark and $r$ is the transverse separation between the quark and the antiquark at equal ordinary time}, $L$ is the orbital quantum number, $M$ is the mass of the meson and $\Phi(\zeta)$ is the transverse mode of the light-front wavefunction which is itself given by 
\begin{equation}
\phi(x,\zeta, \varphi)=\frac{\Phi(\zeta)}{\sqrt{2\pi \zeta}} f(x) \mathrm{e}^{i L \varphi} \;.
\label{factorized-lc}
\end{equation}

It remains a challenge to derive the confining potential $U(\zeta)$ from first-principles QCD but after identifying $\zeta$ with
the co-ordinate in the fifth dimension and angular momentum with the fifth dimensional mass\footnote{$(m_5 R)^2=-(2-J)^2 + L^2$ where $R$ is the radius of curvature in AdS space.}, 
equation~\eqref{LFeigenvalue}
describes the propagation of spin-$J$ string modes, in which
case $U(\zeta)$ is determined by the choice for the dilaton
field. Remarkably, it can be shown \cite{Brodsky:2013npa}  that the dilaton profile is constrained to be quadratic so that the resulting confining potential is given by 
\begin{equation}
 U(\zeta)=\kappa^4 \zeta^2 + 2\kappa^2(J-1) \;.
\label{quadratic-dilaton}
\end{equation}
 Solving equation \eqref{LFeigenvalue} with this confining potential yields the eigenfunctions
\begin{equation}
\Phi_{nL}(\zeta)=\kappa^{1+L}\sqrt{\frac{2n\!}{(n+L)\!}} \zeta^{1/2+L} \exp\left(-\kappa^2 \zeta^2/2\right) L_n^L(\kappa^2 \zeta^2)
\label{heigenfunctions}
\end{equation}
with the corresponding eigenvalues
\begin{equation}
M_{nL,S}^2=4\kappa^2\left(n+L+\frac{S}{2}\right) \;.
\label{eigenvalues}
\end{equation}

\section{The $\rho$ meson wavefunction}
For the $\rho$ meson, $n=1$, $L=0$ and $J=1$ so that $\kappa=M_{\rho}/\sqrt{2}=0.54$ GeV. In equation \eqref{factorized-lc}, $f(x)$ is fixed by comparing the expressions for the pion EM form factor in light-front QCD and in AdS space. This yields $f(x)=\sqrt{x(1-x)}$ so that the resulting  
AdS/QCD for the $\rho$ is then given by
\begin{equation}
 \phi(x,\zeta) \propto \sqrt{x(1-x)} \exp \left(-\frac{\kappa^2 \zeta^2}{2}\right) \exp\left(-\frac{m_f^2}{2\kappa^2 x (1-x)} \right)
\label{lcwf-massive-quarks}
\end{equation}
where the dependence on the quark mass has been introduced according to the prescription by Brodsky and de T\'eramond \cite{Brodsky:2008kp}. Here  we use a light quark mass $m_f=0,14$ GeV \cite{Forshaw:2012im}. 

An earlier procedure to obtain the meson wavefunction is by boosting a non relativistic gaussian Schroedinger wavefunction \cite{Nemchik:1996cw,Forshaw:2003ki} which results in the so-called 
Boosted Gaussian (BG): 
\begin{equation}
\phi^{{\mathrm{BG}}} (x,\zeta) \propto x(1-x) \;
\exp \left(\frac{m_f^{2}R^{2}}{2}\right)
\exp \left(-\frac{m_f^{2}R^{2}}{8 x(1-x)}\right) \; \exp \left(-\frac{2 \zeta^{2}}{{R}^{2}}\right) \;.
\label{original-boosted-gaussian} 
\end{equation}
If $R^2=4/\kappa^2$ then
the two wavefunctions differ only by a factor of $\sqrt{x(1-x)}$,
which is not surprising given that
in both cases confinement is modelled by a harmonic oscillator \cite{Forshaw:2012im}. In what follows we shall consider a parameterization that accommodates
both the AdS/QCD and the BG wavefunctions: 
\begin{equation}
 \phi(x,\zeta) \propto [x(1-x)]^\beta \exp \left(-\frac{\kappa^2 \zeta^2}{2}\right) \exp\left(-\frac{m_f^2}{2\kappa^2 x (1-x)} \right)~.
\label{lcwf-massive-quarks-fit}
\end{equation}
The AdS/QCD wavefunction is obtained by fixing $\beta=0.5$ and $\kappa=0.55$ GeV where as the BG wavefunction is obtained by fixing $\beta=1$ and treating $\kappa$ as a free parameter.

\section{Results and conclusions}
To compute the rate for diffractive $\rho$ production, we use the dipole model of high-energy scattering \cite{Nikolaev:1990ja,Nikolaev:1991et,Mueller:1993rr,Mueller:1994jq} in which the scattering amplitude for diffractive $\rho$ meson production is a convolution of the photon
and vector meson $q\bar{q}$ light-front wavefunctions with the total cross-section
to scatter a $q\bar{q}$ dipole off a proton. QED is used to determine
the photon wavefunction and the dipole cross-section can be extracted from the precise data on the deep-inelastic structure
function $F_2$ \cite{Soyez:2007kg,Forshaw:2004vv}. This formalism can then be used to predict rates for vector meson production and diffractive DIS \cite{Forshaw:2003ki,Forshaw:2006np} or to
to extract information on the $\rho$ meson wavefunction using the HERA data on diffractive $\rho$ production \cite{Forshaw:2010py,Forshaw:2011yj}. Here we use it to test whether the HERA data prefer the AdS/QCD wavefunction given by equation \eqref{lcwf-massive-quarks}. To do so, we compute the $\chi^2$ per data point in the $(\beta, \kappa)$ parameter space using the parametrization \eqref{lcwf-massive-quarks-fit} for the $\rho$ wavefunction.\footnote{We include the electroproduction data and decay width datum in the fit.}  

\begin{figure}
 \includegraphics[height=.3\textheight]{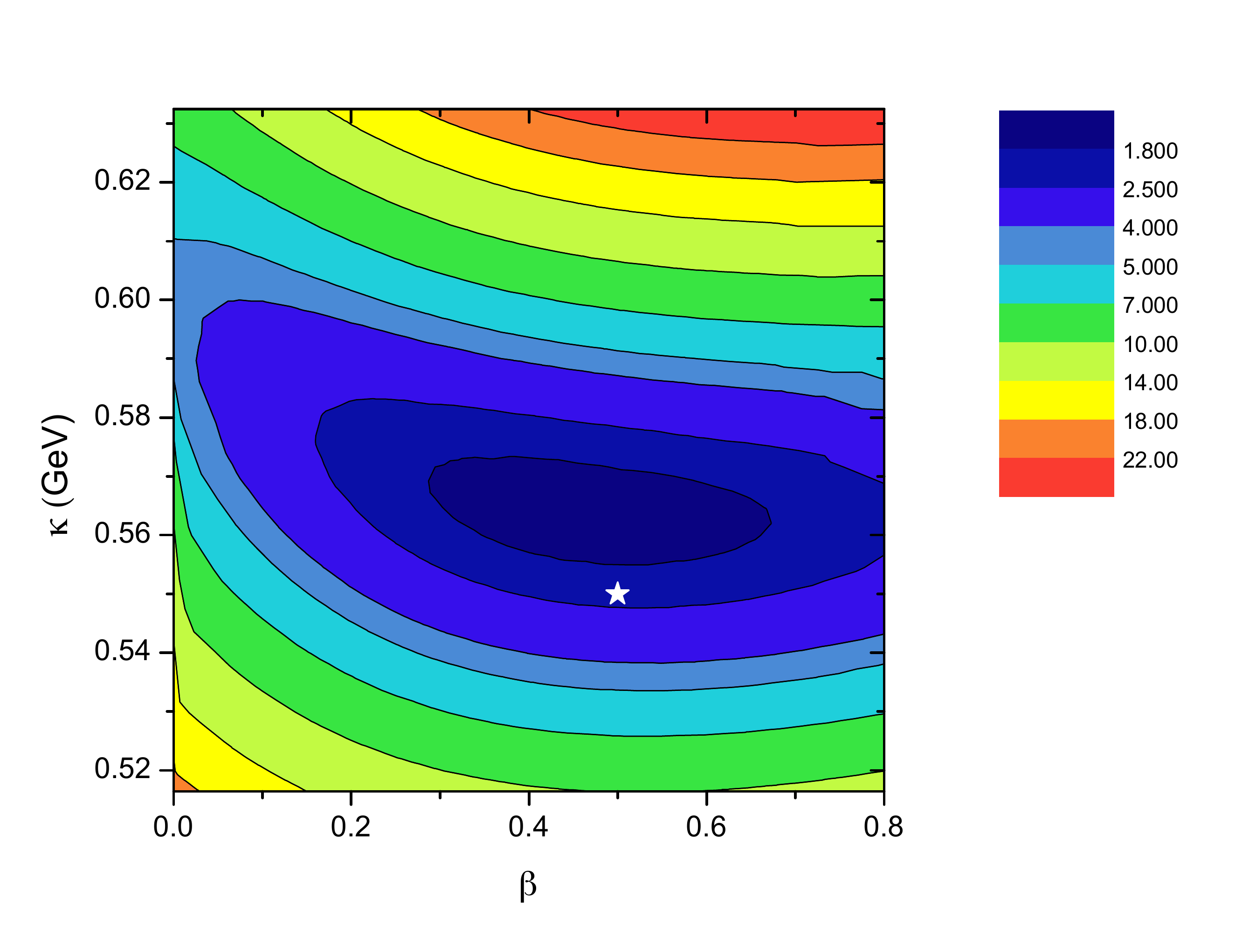}
  \caption{The $\chi^2$ distribution in the $(\beta,\kappa)$ parameter space.  The AdS/QCD prediction is the white star.}
\label{fig:contour}
\end{figure}

Figure \ref{fig:contour} confirms that the AdS/QCD
prediction lies impressively close to the minimum in $\chi^2$. The best fit has a $\chi^2$ per data point equal to $114/76$ and is
achieved with $\kappa=0.56$ GeV and $\beta=0.47$ which should be compared with the AdS/QCD prediction: $\kappa=0.54$ and $\beta=0.5$ shown as the white star on figure \ref{fig:contour}. 
Note that the BG prediction i.e. $\beta=1, \forall \kappa$,  is clearly further away from the minimum in $\chi^2$. 

Finally, we note that these results are produced using a particular Color Glass Condensate dipole model \cite{Soyez:2007kg} but that similar results are obtained by using other forward dipole models \cite{Forshaw:2004vv}  that fit the $F_2$ structure function data. It remains to be seen how the $\chi^2$ distribution changes if a more sophisticated dipole model, such as the recent impact parameter saturation model \cite{Rezaeian:2012ji} which fits the combined HERA $F_2$ data, is used. 

\bibliographystyle{JHEP}
\bibliography{RSandapenJRF}

\end{document}